\documentclass[dvips]{article}

\usepackage{icrc2011}

\title{Observation of selected IBLs and LBLs with VERITAS}
\newcommand{\etal}{\MakeLowercase{\textit{et al. }}} % "et al."
\shorttitle{P. Majumdar\etal Observations of LBLs/IBLs}

\authors{Pratik Majumdar$^{1}$ for the VERITAS Collaboration$^{2}$}
\afiliations{$^1$University of California, Los Angeles\\
$^{2}$~see J.Holder et al, these proceedings or http://veritas.sao.arizona.edu/conferences/authors?icrc2011
}
\email{pratik@astro.ucla.edu}

\abstract{The improved sensitivity of current-generation atmospheric-Cherenkov telescope (ACT) arrays enables us 
to probe for the first time low-frequency-peaked and intermediate-frequency-peaked BL Lac (LBL and IBL, respectively) 
objects as very high energy (VHE) $\gamma$-ray emitters. VERITAS is currently one of the most sensitive VHE gamma-ray 
observatories in the world and is particularly suited to study these objects. In this paper we will present recent 
results from VERITAS observations of a selected set of LBLs and IBLs. 
}
\keywords{gamma rays, active galactic nuclei}

\begin{document}
\maketitle

%Begin the section.
\section{Introduction}
\vspace{-0.4cm}
Blazars are a class of radio-loud active galactic nuclei (AGN) that have relativistic
jets oriented towards the Earth. Observationally blazars show broadband non-thermal emission
over a wide range of frequencies, from IR (flat-spectrum radio quasars (FSRQs) and LBL) to optical/UV (IBL) and 
X-rays. VHE ($E >$ 100 GeV) gamma rays have been detected from a large number of blazars 
by ground-based atmospheric-Cherenkov telescopes (ACTs). However most of the 
confirmed TeV blazars are high-frequency BL Lac objects (HBLs), as opposed to quasars and LBL/IBLs
that consitute the majority of the population in the MeV-GeV band. It is only recently that improved 
senstitivities of current generation ACTs have allowed us for the first time to probe LBL and IBLs as VHE
emitters. Here we report on the observations of selected LBLs and IBLs, namely W Comae, 3C~66A, PKS~1424+24, 
1ES1440+122 and 1ES1215+303 by VERITAS~\cite{holder2011}.
and discuss the implication of these observations in the context of 
multiwavelength campaigns.     

\vspace{-0.4cm}
\section{VERITAS Array and Data Analysis}
\vspace{-0.4cm}
VERITAS is an array of four imaging Cherenkov telescopes located at the Fred Lawrence 
Whipple Observatory in southern Arizona at a height of 1268 $m$ a.s.l. It combines 
a large effective area ( $\sim$ 10$^{5}$ m$^{2}$ ) over a wide energy range (100 GeV to 
30 TeV) with good energy resolution (15-25\%) and high angular resolution (R$_{68\%}$ $<$ 0.1$^{\circ}$).
The field of view of VERITAS is 3.5$^{\circ}$. The high sensitivity of VERITAS allows the detection
of sources with a flux of 1\% of the Crab Nebula flux in about 25 hours. For more details on the VERITAS
instrument and analysis, please refer to \cite{holder2006, weekes2002, acciari2008a}. 

%The data on various sources are taken in wobble mode, wherein the source is positioned
%at a fixed offset of 0.5$^{\circ}$ from the camera center. This allows a simultaneous estimate of the 
%background. The analysis steps consist of calibration and integration of the flash-ADC traces, image
%cleaning, second-moment parametrization of images in each telescope, stereoscopic reconstruction of the 
%event impact position and direction, gamma-hadron separation and generation of photon maps. Far more 
%numerous background events are rejected by comparing the shape parameters of the event images 
%(namely mean scaled width and mean scaled lengthi~\cite{acciari2008a}) with the 
%expected shapes of gamma-ray showers modeled by Monte Carlo simulations. An additional cut on the arrival
%direction of incoming gamma rays reject more than 99.9\% of the background. The energy of each event is 
%estimated from detailed Monte Carlo simulations of extensive air showers and of the response of the telescopes, 
%focal plane detectors and electronics. The systematic error in the energy estimation is dominated by 
%uncertainties in the atmospheric models used in the simulations and various detector parameters.    

\vspace{-0.4cm}
\section{Observations of IBLs}
\vspace{-0.4cm}
\subsection{W Comae}

 \begin{figure}[!t]
  \vspace{5mm}
  \centering
  \includegraphics[width=3.2in]{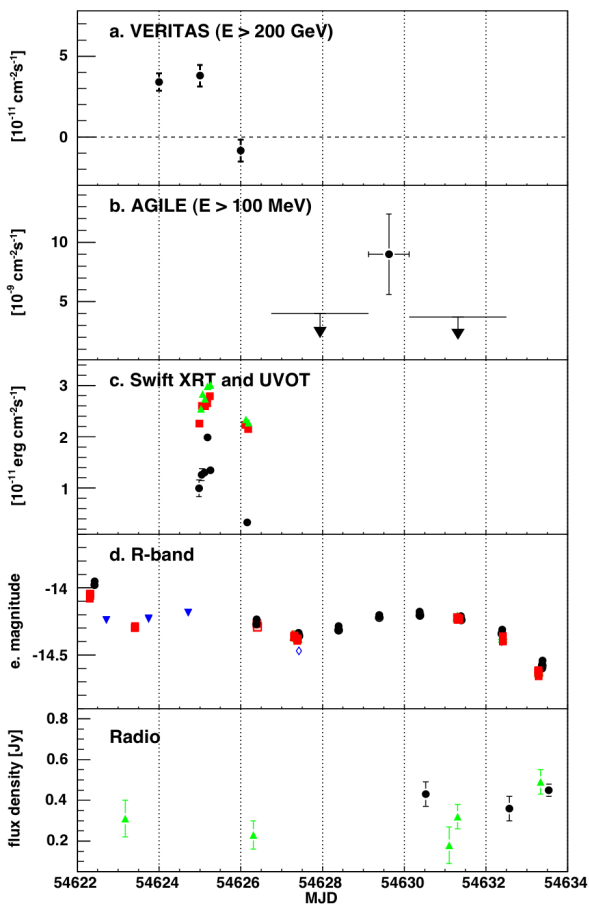}
  \vspace{-0.4cm}
  \caption{\small Multiwavelength light curve of W Comae for MJD 54622 to 54634. Panel a: VHE light curve
by VERITAS (E $>$ 200 GeV), b: light curve by AGILE (E $>$ 100 MeV), c: X-rays({\it Swift} and XMM-Newton) and
{\it Swift} UVOT, d: R-band optical lightcurve from Tuorla, Abastumani, San Pedro Martir, Sapienza Univ., KVA,
Crimean and Torino, e(lowerpost panel): radio light curves from UMRAO, Mets\"ahovi. 
See~\cite{reyes2009} for more details }
  \label{wcom_mwl}
 \end{figure}

\begin{figure}[!t]
  \vspace{5mm}
  \centering
  \includegraphics[width=4.0in]{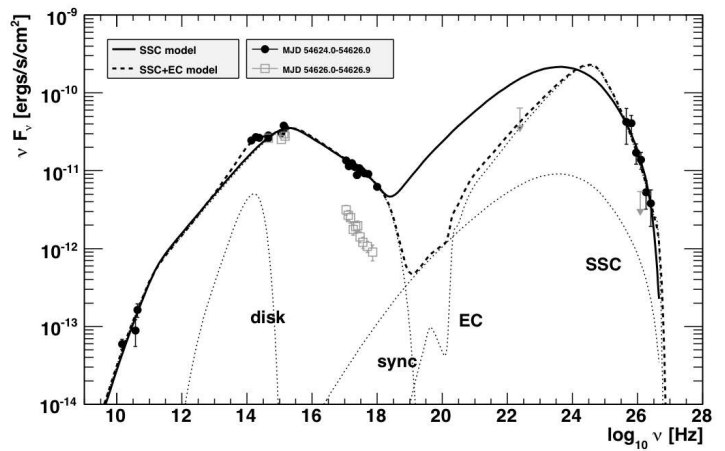}
  \vspace{-0.4cm}
  \caption{\small Spectral energy distribution of W Comae for the flare period including VERITAS, {\it Swift} 
XRT/UVOT, optical 
and radio data. Results from SSC and SSC+EC models are shown as continuous and dashed lines respectively. The different
components of the SSC+EC model are indicated by dotted lines.
See~\cite{reyes2009} for more details.} 
  \label{wcomSED}
 \end{figure}

\vspace{-0.4cm}
W Comae (ON 231, {\it z} =0.102) was the first IBL to be detected at VHE~\cite{acciari2008b}. 
It was discovered as a VHE source by VERITAS during observations carried out over a 4-month period
(Jan 2008 to Apr 2008). During this time, a strong gamma-ray outburst was measured over a 
period of 4 days, when the source went into a flaring state in the middle of March 
reaching 9\% of Crab Nebula flux. The photon index during the flare was measured to be  
$\Gamma$ = 3.81 $\pm$ 0.35$_{stat}$ $\pm$ 0.34$_{sys}$.
In June 2008, the source 
again went to a high state and was approximately three times brighter than during 2008 March observations, 
triggering a multiwavelength campaign which included AGILE, {\it Swift} and XMM-Newton. VERITAS observed 
W Comae for 230 minutes in between June 7 and 9. The average flux on June 7-8 was 2.5-3 times higher than that
observed during March 2008. The differential photon flux can be fitted with a power law with an index 
$\Gamma$ = 3.68 $\pm$ 0.22$_{stat}$ $\pm$ 0.3$_{sys}$.
Figure~\ref{wcom_mwl} shows the 
multiwavelength light curve of W Comae (MJD 54622 to 54634)~\cite{reyes2009}. 
The initial detection of the flare by VERITAS was followed by 
observations in high-energy gamma rays (AGILE, E$_{\gamma}$ $>$ 100 MeV), X-rays ({\it Swift} and XMM-Newton), 
UV and ground based optical and radio monitoring instruments through the GASP-WEBT consortium. 
The measurements revealed strong variability in both gamma rays and X-rays on timescales of days.
The broadband spectral 
energy distribution (SED) during the flare (MJD 54624 to 54626) has been modeled using a leptonic 
one-zone jet model~\cite{bottchiang2002}. 
The SED (see Figure~\ref{wcomSED}) can be modeled by a simple leptonic synchrotron self-Compton model (SSC), however the wide 
separation of the peaks in the SED requires a rather low ratio of magnetic field to electron energy density. 
On the other hand, if one invokes external-Compton emission, magnetic field parameters close to 
equipartition can be obtained.   

\vspace{-0.4cm}
\subsection{PKS~1424+240}
\begin{figure}[!t]
  \vspace{5mm}
  \centering
  \includegraphics[width=3.8in]{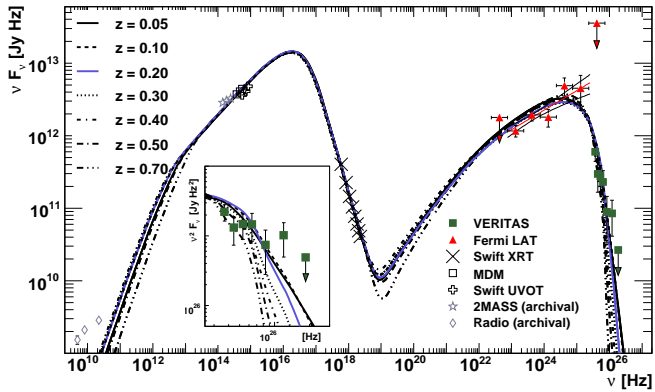}
  \vspace{-0.4cm}
  \caption{\small Spectral energy distribution of PKS~1424+240. The lines show the SSC model fits for different
redshifts. The inset shows a zoom of the SED on VERITAS data in $\nu^{2}F_{\nu}$ representation. See~\cite{otte2009}
for more details} 
  \label{pkssed}
 \vspace{-0.3cm}
 \end{figure}

\vspace{-0.4cm}
PKS~1424+240 was detected as a radio source and was classified as a blazar from optical
polarisation studies~\cite{impey}. Non-thermal X-rays were later reported from this source~\cite{fleming1993}. 
The redshift of PKS~1424+240 is unknown. A lower limit on the redshift of {\it z} $>$ 0.06 was measured by 
\cite{scarpa} and limit of {\it z} $<$ 0.67 was obtained by~\cite{sbarufatti2005}, both assuming a 
minimum luminosity of the host galaxy. PKS~1424+240 was observed by VERITAS for about 37.3 hours between 2009 
February 19 and June 21, 65\% of the data having been taken with 3 telescopes. The analysis of the data resulted 
in the discovery of PKS~1424+240. The statistical significance of the observed excess 
of gamma-ray events is 8.5 standard deviations~\cite{otte2009}. The photon spectrum above 140 GeV measured by VERITAS 
is well described by a power law with an index of 3.8 $\pm$ 0.5$_{stat}$ $\pm$ 0.3$_{sys}$ and a flux 
normalisation at 200 GeV of (5.1 $\pm$ 0.9$_{stat}$ $\pm$ 0.5$_{sys}$) $\times$ 10$^{-11}$ cm$^{-2}$s$^{-1}$TeV$^{-1}$. 
No varability is observed in the VHE flux over the period of observations. Flux variability is also not 
observed in the contemporaneous high energy observations with {\it Fermi}~ Large Area Telescope (LAT). Contemporaneous X-ray
and optical data were also obtained from {\it Swift} XRT and MDM observatory. The broadband SED can be well described by 
a one-zone SSC model~\cite{bottchiang2002} if one assumes a redshift of $<$ 0.1. 
Figure~\ref{pkssed} shows the SED of PKS~1424+240. The inset
of Figure~\ref{pkssed} illustrates that above {\it z} $\sim$ 0.2, the model VHE gamma ray spectrum becomes too steep 
compared to observed VERITAS spectrum and the model also requires unreasonably large Doppler factors. 
An attempt to improve the fit in the gamma-ray band by including an EC component results in a steeper VHE gamma-ray 
spectrum. An upper limit on the redshift of the source can be obtained using the photon index as measured with 
{\it Fermi} in combination with recent extragalactic background light absorption models~\cite{frances2008, finke2009} and 
the VHE spectrum measured by VERITAS. Using this method the upper limit on the redshift of the source has been estimated to be 0.66. 

\vspace{-0.4cm}
\subsection{3C~66A}
\begin{figure}[!t]
  \vspace{5mm}
  \centering
  \includegraphics[width=3.1in]{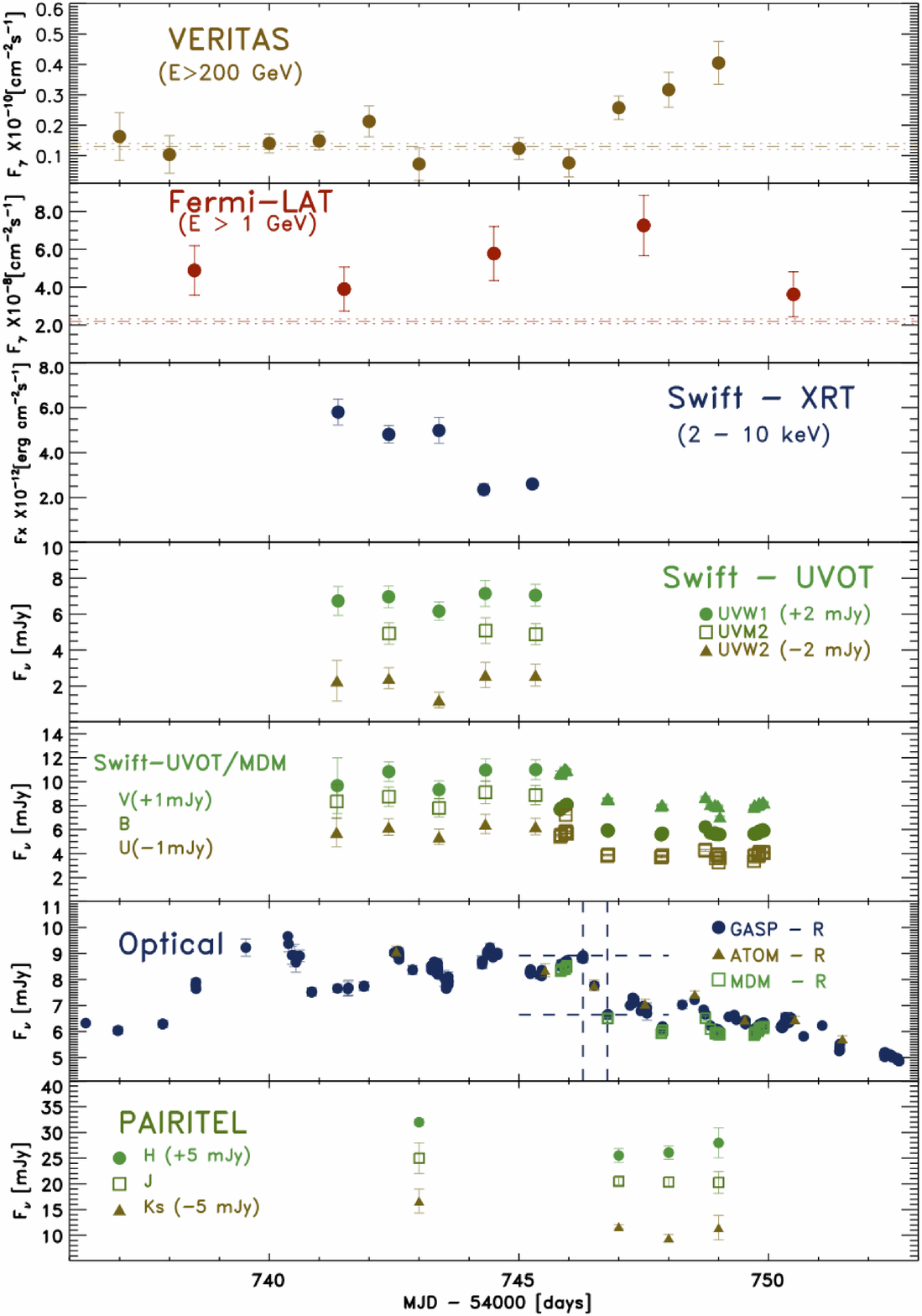}
  \vspace{-0.4cm}
  \caption{\small 3C~66A light curves covering the period centered on the gamma-ray flare (2008 October 1-10) 
The panels include observations made with VERITAS, {\it Fermi}-LAT, {\it Swift} XRT/UVOT, MDM, GASP-WEBT, ATOM and PAIRITEL. 
See~\cite{reyes2010} for details. }
  \label{3c66a_flare_mwl}
 \end{figure}

\begin{figure}[!t]
  \vspace{5mm}
  \centering
  \includegraphics[width=3.3in]{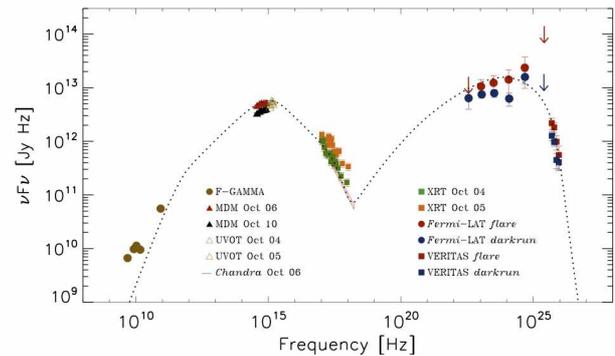}
  \vspace{-0.5cm}
  \caption{\small Broadband SED of 3C~66A during the October 2008 multiwavelength campaign. 
See~\cite{reyes2010} for details.}
  \label{3c66a_sed}
 \vspace{-0.3cm}
 \end{figure}

\vspace{-0.4cm}
3C~66A was first identified as a quasi-stellar object (QSO) using optical
observations~\cite{Wills:1974eu}. It was subsequently classified as a BL Lac object based
on its significant optical and X-ray variability
\cite{Maccagni:1987qe}. VERITAS observed 3C\,66A for 14 hours from September 2007 through
January 2008 (hereafter, the 2007-2008 season). Later, from September
through November 2008 (hereafter, the 2008-2009 season), a further 46
hours of data were taken. After removing data with poor weather conditions or hardware-related 
problems, 4.7 hours live time in 2007-2008 and 28.1 hours in 2008-2009 were used for final analysis.
Analysis of the data resulted in an excess of 1791 events from the direction of 3C\,66A.
The excess corresponds to a statistical
significance of 21.1 $\sigma$~\cite{perkins2008}. 
Here we report on the detection of a strong flare in October 2008 with night-by-night 
VHE flux variability~\cite{reyes2010}. Contemporaneous variability was also detected from optical to infra-red 
wavelengths and also in the {\it Fermi}-LAT energy band. Follow-up observations were carried out in radio
(Medicina, Mets\"ahovi, Noto and UMRAO observatories), 
optical (GASP-WEBT, MDM, ATOM), infra-red (PAIRITEL) and X-ray ({\it Swift} XRT, Chandra) wavelengths in order to measure the flux 
and spectral variability of the source and obtain a quasi simultaneous SED. 
The average flux in VHE gamma rays above 200 GeV for the period in October 2008 (MJD 54734-54749) 
is found to be (1.4 $\pm$ 0.2) $\times$ 10$^{-11}$ cm$^{-2}$s$^{-1}$ whereas the flux during the flare period 
(MJD 54747-54749) was found to be (2.5 $\pm$ 0.4) $\times$ 10$^{-11}$ cm$^{-2}$s$^{-1}$.  
Figure~\ref{3c66a_flare_mwl} shows the multiwavelength light curve of 3C~66A centred around the flare 
(October 1 to October 10). 
The light curves obtained show strong variability at every observed wavelength and in particular the flux increase 
seen by VERITAS and {\it Fermi}-LAT coincides with an optical outburst. The overall SED (see Figure~\ref{3c66a_sed})
can be reasonably modeled using a pure
SSC model, or an SSC+EC model~\cite{bottchiang2002} with an external near infra-red radiation field as an 
additional source for Compton scattering. 
However the pure SSC model requires a large emission region which is inconsistent with the intra-night variability in optical 
wavelengths, and very low magnetic fields. Again, a smaller emission region would require unreasonably large Doppler factors,
difficult to reconcile within the model.
In contrast, the SSC+EC model can account for variability timescales of a few 
hours and predict a more reasonable magnetic field even though it is below equipartition. 

\vspace{-0.6cm}
\subsection{1ES1440+122}
\vspace{-0.4cm}
1ES1440+122 was initially identified as an AGN in the Einstein Slew Survey~\cite{elvis1992}. It is
surrounded by $\sim$ 20 galaxies, which suggests that it may belong to a cluster of galaxies~\cite{heidt1999}.
1ES1440+122 is classified as an intermediate-frequency-peaked BL Lac object by~\cite{nieppola2006} 
and its redshift, {\it z} = 0.162, is based on the identification of at least 
three spectral lines~\cite{sbarufatti2006}. 
1ES1440+122 belongs to the the first {\it Fermi}-LAT MeV-GeV catalog 
~\cite{abdo2010}, having an average photon index of 1.77$\pm$0.25. It is one of the hardest spectrum sources
(top 10\%) in the catalog. The blazar was observed with the VERITAS 
array for $\sim$ 50 hours between June 2008 and July 2010. Analysis of these data has resulted 
in the detection of the object with a statistical significance of 5.5$\sigma$. The observed flux 
is less than 1\% of the flux from the Crab Nebula above 200 GeV and the observed VHE light curve does not 
show any evidence for variability~\cite{ong2010}.

\vspace{-0.4cm}
\subsection{1ES~1215+303}
\vspace{-0.4cm}
1ES 1215+303, also commonly referred to as ON 325, was first detected in the B2 408 MHz survey
conducted with the Bologna Northern Cross telescope.
It was detected at X-ray energies by the Einstein observatory. 1ES 1215+303 has been classified as a bright
IBL based on a synchrotron peak located at 10$^{15.6}$ Hz~\cite{nieppola2006}. The redshift of the object is listed
as {\it z} = 0.130 in NED but is uncertain, according to White et al. it is 0.237~\cite{white2000}.
1ES 1215+303 was first detected
in the VHE band during a brief flare over a period of 4-nights by MAGIC~\cite{magic2011} in early January 2011.
The VHE flux claimed during this flare was 2.0\% of the Crab Nebula flux above 250 GeV. The field of view of 1215+303
also contains 2 other VHE sources, the well-known BL Lac object 1ES~1218+304 and another IBL W~Comae~\cite{benbow2011}.
VERITAS has been regularly observing 1ES~1218+304 since all
4 telescopes became operational. Thus the data set spans from 2008 to 2011. However, since the east wobble position
is 1.3$^\circ$ away, it was not used for the analysis of data of 1ES 1215+303. On the other hand, the west wobble position
is very close to the source, so this required a special analysis, namely the ring background model~\cite{acciari2008a}.
Approximately 55 hrs of data have been logged and a standard analysis yields an excess of about 235 events
and a significance of 6.1$\sigma$. The flux corresponds to $\sim$ 1\% of the Crab Nebula flux.

\vspace{-0.4cm}
\subsection{BL Lacertae}
\begin{figure}[!t]
  \vspace{5mm}
  \centering
  \includegraphics[width=2.75in]{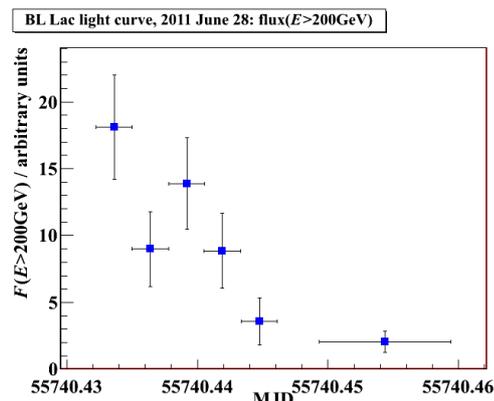}
  \vspace{-0.4cm}
  \caption{\small Light curve of BL Lac on the night of the flare. The first 5 data points are 4-min time bins and the
last data point is a 20-min bin. The flux changes by a factor $\sim$ 2 in 4 minutes.}
  \label{bllac_lightcurve}
 \vspace{-0.3cm}
 \end{figure}

\vspace{-0.4cm}
BL Lacertae(1ES~2200+420,{\it z}=0.069) is the archetype BL Lac object, and belongs
to the sub-class of low-frequency-peaked BL Lac (LBL) objects. 
It was discovered at VHE by MAGIC during a flare in 2005 with a flux of
$\sim$ 3\% of the Crab Nebula flux above 200 GeV. No variability was recorded in the data-set. No significant 
excess was found in the MAGIC follow-up observations in 2006. The source has been monitored by VERITAS in 2010 and 
an upper limit of $<$ 3\% of Crab Nebula flux has been derived from these observations. In May 2011, several observatories
reported an increase in activity of the source. Following this, VERITAS monitoring observations commenced on
26$^{th}$ May and continued on several nights through June. On June 28, a flare was observed with VERITAS during a 
40-minute observation under moonlight conditions~\cite{atel2011}. 
Twilight stopped any further observations on that night. The integral 
flux reached as high as 50\% of Crab Nebula flux above 300 GeV (20~$\sigma$ significance) during the first 20-minute run 
and decreased to $\sim$ 10\% of Crab Nebula flux (5~$\sigma$ significance) in the next 20 minutes. Figure~\ref{bllac_lightcurve} 
shows the light curve
of the source on the night of the flare (see details in caption). The VHE spectrum is soft with an index of about 
$\Gamma$=-3.4$\pm$0.4. No clear signal was observed during 7 hours of 2011 VERITAS monitoring observations 
preceeding the night of the VHE flare. 
    
{\noindent \bf \small Acknowledgements}
{\small This research is supported by grants from the US Department of Energy, the US National Science Foundation, and the
Smithsonian Institution, by NSERC in Canada, by Science Foundation Ireland, and by STFC in the UK. We acknowledge the
excellent work of the technical support staff at the FLWO and the collaborating institutions in the construction and
operation of the instrument. }

%\vspace{\baselineskip}

\clearpage


\vspace{-0.5cm}
\begin{thebibliography}{}
\vspace{-0.4cm}
\footnotesize{
\bibitem{holder2011} Holder, J. et al (VERITAS Collaboration), these proceedings 

\bibitem{holder2006} Holder, J. et al (VERITAS Collaboration), Astroparticle Phys, 2006, {\bf 25}:391

\bibitem{weekes2002} Weekes, T.C. et. al., Astroparticle Phys, 2002, {\bf 17}:221

\bibitem{acciari2008a} Acciari, V.A. et al (VERITAS Collaboration), ApJ, 2008, {\bf 679}:1427

\bibitem{acciari2008b} Acciari, V.A. et al (VERITAS Collaboration), ApJ, 2008, {\bf 684}:L73

\bibitem{bottchiang2002} B\"ottcher, M. \& Chiang, J. ApJ, 2002, {\bf 581}:127

\bibitem{reyes2009} Acciari, V.A. et al (VERITAS Collaboration), ApJ, 2009, {\bf 707}:612

\bibitem{impey} Impey, C.D. \& Tapia, S. ApJ, 1988, {\bf 333}:666

\bibitem{fleming1993} Fleming, T.A. AJ, 1993, {\bf 106}, 1729

\bibitem{scarpa} Scarpa, R. \& Falomo, R. A\&A, 1995, {\bf 303}:656

\bibitem{sbarufatti2005} Sbarufatti, B. et al, ApJ, 2005, {\bf 635}:173

\bibitem{otte2009} Acciari, V.A. et al (VERITAS Collaboration), ApJ, 2010, {\bf 708}:L100

\bibitem{frances2008} Franceschini, A. et al, A\&A, 2008, {\bf 487}:837

\bibitem{finke2009} Finke, J.D. et al, 2009, arXiv:0905.1115

\bibitem{Wills:1974eu} Wills, B.J. \& Wills, D. ApJ, 1974, {\bf 190}:L97

\bibitem{Maccagni:1987qe} Maccagni, D. et al, A\&A, 1987, {\bf 178}:21

\bibitem{perkins2008} Acciari, V.A. et al (VERITAS Collaboration), ApJ, 2009, {\bf 693}:L104

\bibitem{reyes2010} Abdo, A.A. et al, ApJ, 2011, {\bf 726}:43

\bibitem{elvis1992} Elvis, M et al, ApJS, 1992, {\bf 80}:257

\bibitem{heidt1999} Heidt, J. et al, A\&A, 1999, {\bf 341}:683

\bibitem{nieppola2006} Nieppola, E. et al, A\&A, 2006, {\bf 445}:451

\bibitem{sbarufatti2006} Sbarufatti, B. et al, A\&A, 2006, {\bf 457}:35

\bibitem{abdo2010} Abdo, A.A. et al, ApJS, 2010, {\bf 188}:405

\bibitem{ong2010} Ong, R. (VERITAS Collaboration), ATEL\#2786
 
\bibitem{white2000} White, R.L. et al, ApJS, 2000, {\bf 126}:133

\bibitem{benbow2011} Benbow, W., these proceedings

\bibitem{magic2011} Mariotti, M. (MAGIC Collaboration), ATEL\#3100

\bibitem{atel2011} Ong, R. (VERITAS Collaboration), ATEL\#3459

}
\end{thebibliography}
\end{document}